\documentclass[12pt]{article}

\usepackage{epsfig}
\usepackage{geometry}
\usepackage{caption}

\def\balpha{\mbox{\boldmath $\alpha$}}

\usepackage{amssymb}

\def\br{{\rm BR}}
\def\Be{{\rm Be}}

\begin{document}

\begin{titlepage}

\baselineskip 24pt

\begin{center}

{\Large {\bf Accommodating three low-scale anomalies\\ ($g -2$, Lamb shift, and 
Atomki) in the\\ framed standard model}}

\vspace{.5cm}

\baselineskip 14pt

{\large Jos\'e BORDES \footnote{Work supported in part by Spanish
    MINECO under grant FPA2017-84543-P,
Severo Ochoa Excellence Program under grant SEV-2014-0398. }}\\
jose.m.bordes\,@\,uv.es \\
{\it Departament Fisica Teorica and IFIC, Centro Mixto CSIC, Universitat de 
Valencia, Calle Dr. Moliner 50, E-46100 Burjassot (Valencia), 
Spain}\\
\vspace{.2cm}
{\large CHAN Hong-Mo}\\
hong-mo.chan\,@\,stfc.ac.uk \\
{\it Rutherford Appleton Laboratory,\\
  Chilton, Didcot, Oxon, OX11 0QX, United Kingdom}\\
\vspace{.2cm}
{\large TSOU Sheung Tsun}\\
tsou\,@\,maths.ox.ac.uk\\
{\it Mathematical Institute, University of Oxford,\\
Radcliffe Observatory Quarter, Woodstock Road, \\
Oxford, OX2 6GG, United Kingdom}

\end{center}

\vspace{.3cm}

\begin{abstract}

The framed standard model (FSM) predicts a $0^+$ boson with mass around 20 MeV 
in the ``hidden sector'', which mixes at tree level with the standard Higgs 
$h_W$ and hence acquires small couplings to quarks and leptons which can be 
calculated in the FSM apart from the mixing parameter $\rho_{Uh}$.  The 
exchange of this mixed state $U$ will contribute to $g - 2$ and to the Lamb 
shift.  By adjusting $\rho_{Uh}$ alone, it is found that the FSM can satisfy 
all present experimental bounds on the $g - 2$ and Lamb shift anomalies for 
$\mu$ and $e$, and for the latter for both hydrogen and deuterium.

The FSM predicts also a $1^-$ boson in the ``hidden sector'' with a mass of 17 
MeV, that is,  right on top of the Atomki anomaly $X$.  This mixes with the  
photon at 1-loop level and couples thereby like a dark photon to quarks and 
leptons. It is however a compound state and is thought likely to possess 
additional compound couplings to hadrons.  By adjusting the mixing parameter 
and the $X$'s compound coupling to nucleons, the FSM can reproduce the 
production rate of the $X$ in beryllium decay as well as satisfy all the bounds 
on $X$ listed so far in the literature.

The above two results are consistent in that the $U$, being $0^+$, does not 
contribute to  the Atomki anomaly if parity and angular momentum are conserved, 
while $X$, though contributing to $g - 2$ and Lamb shift, has smaller couplings 
than $U$ and can, at first instance, be neglected there.

Thus, despite the tentative nature of the 3 anomalies in experiment on the one 
hand and of the FSM as theory on the other, the accommodation of the former in 
the latter has strengthened the credibility of both.  Indeed, if this FSM 
interpretation were correct, it would change the whole aspect of the anomalies
from just curiosities to windows into a vast hitherto hidden sector comprising
at least in part the dark matter which makes up the bulk of our universe. 

\end{abstract}

\end{titlepage}

\clearpage

\section{Introduction}

Three deviations from the standard model (SM) recently observed in experiment, 
though not yet fully established, have attracted much theoretical attention for 
suggesting new physics.
\begin{itemize}
\item {\bf [A]} First, the accurate measurement \cite{g-2exptmuon}
  of the magnetic moment 
  $g - 2$ for the muon shows \cite{g-2-analyses1,g-2-analyses2} an up to
  $4\ \sigma$ deviation from SM prediction while $g - 2$ for the electron agrees 
well with the SM prediction \cite{g-2exptelectron}.  For
\begin{equation}
a_{(\mu,e)} = (g_{(\mu,e)} - 2)/2 ,
\label{amue}
\end{equation}
these analyses  give
\begin{eqnarray}
\Delta a_\mu & = & (a_\mu)^{\rm EXP} - (a_\mu)^{\rm SM} = (2.07-3.67) \times 10^{-9}, \\ 
\Delta a_e & = & (a_e)^{\rm EXP} - (a_e)^{\rm SM} < 1.5 \times 10^{-12}.
\label{gminus2}
\end{eqnarray}
\item {\bf [B]} Second, the Lamb shift between the 2S and 2P levels for muonic 
  hydrogen \cite{LSmuonichydrogen1,LSmuonichydrogen2}
  and muonic deuterium \cite{LSmuonicdeuterium1,LSmuonicdeuterium2}
  differ from the SM expected 
values by:
\begin{eqnarray}
\delta E_\mu^H & = & (- 0.363, - 0.251)\ {\rm  meV}, \\ 
\delta E_\mu^D & = & (- 0.475, - 0.337)\  {\rm meV}, 
\label{Lambshift} 
\end{eqnarray} 
while ordinary electronic hydrogen and deuterium show no such
deviations. The 
same anomaly is often phrased as a proton radius puzzle
\cite{protonradius}, ascribing to the
proton an effective radius of $r_p=0.8414$ fm from muonic atoms, in particular
from $\mu H$, as opposed to $r_p=0.877$  fm from $eH$
\cite{LSelectrons1,LSelectrons2} 
and $e-p$ scattering \cite{LSelectronsscattering1,LSelectronsscattering2}.
\item {\bf [C]} Third, the $e^+ e^-$ spectrum from excited ${}^8{\rm Be}^*$ decay 
\cite{Atmokiexpt} shows a $6.8\ \sigma$ bump above the expected
background at the electron-positron invariant mass of
\begin{equation}
m_{e^+ e^-} = 16.70 \pm 0.35\ {\rm (stat)} \pm 0.5\ {\rm (sys)\ MeV},
\label{mAtomki}
\end{equation}
suggestive of a new boson $X$ being produced:
\begin{equation}
{}^8{\rm Be}^* \rightarrow {}^8{\rm Be} \  X, \ \ X \rightarrow e^+ e^-,
\label{Be*decay}
\end{equation}
with 
\begin{equation}
  \frac{\Gamma({}^8{\rm Be}^* \rightarrow {}^8{\rm Be}\ X)}{
    \Gamma ({}^8{\rm Be}^*
    \rightarrow {}^8{\rm Be}\ \gamma)}
   \br (X \rightarrow e^+ e^-) = 5.8 \times 10^{-6}.
\label{brAtomki}
\end{equation}
\end{itemize}
The results {\bf [A]} and {\bf [B]} lack statistical significance while 
{\bf[C]} needs to be independently confirmed.  But if they are real, then the 
consequence is highly significant, being suggestive of new physics outside the 
standard model framework.

It has been suggested that {\bf [A]} and {\bf [B]} are explainable by a new 
scalar boson (say $U$) of low mass while {\bf [C]} points to a new (perhaps 
spin 1) boson (say $X$) of mass around 17 MeV.  The fact that the effects 
{\bf [A]}---{\bf [C]} are all small and no hints of $U$ and $X$ are seen 
anywhere else means that these new bosons must have rather unusual properties
and small couplings to ordinary matter.  The question is then raised as to the 
theoretical origin of these particles $U$ and $X$ which would lie outside the 
standard model framework.  Are they just isolated phenomena or is there a whole 
new class of particles hitherto unknown to us which interact but little with 
the particles we know, and of which both $U$ and $X$ are but examples as tips 
of an iceberg?

\begin{figure}
  \centering
\includegraphics[scale=0.4]{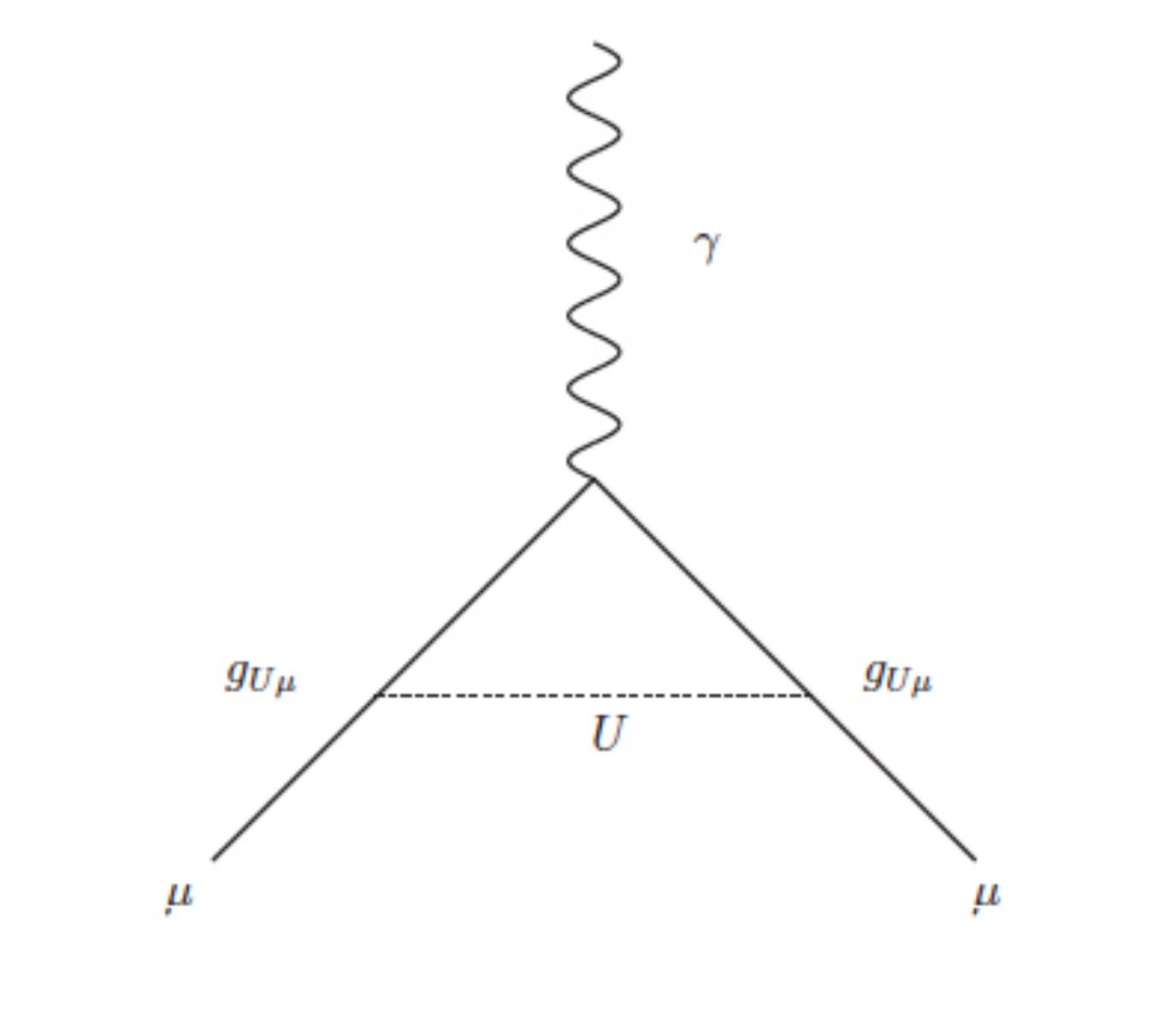}
\caption{Feynman diagram contribution of $U$ to $g-2$ anomaly of $\mu$}
\label{g-2}
\end{figure}

\begin{figure}
  \centering
  \includegraphics[scale=0.3]{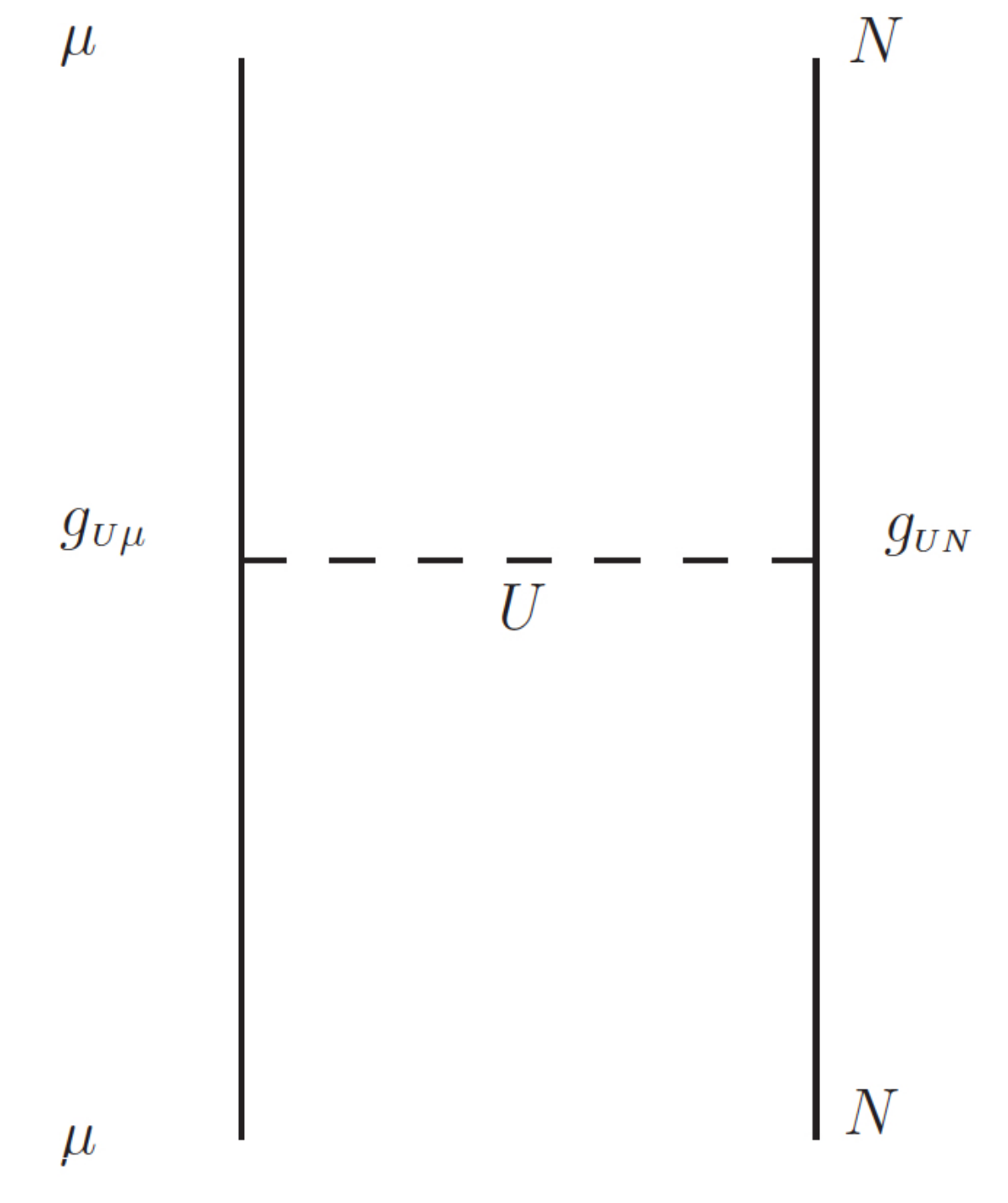}
\caption{Feynman diagram contribution of $U$ to the  Lamb shift
  anomaly of muonic hydrogen and deuterium}
\label{Lambsfig}
\end{figure}

To address these questions in general terms, the framed standard model (FSM) 
seems rather well placed.  It predicts, among other things, a cluster of 
boson states, some scalars $H_{\rm light}$ and some vectors $G_{\rm light}$, all with 
masses around 20 MeV, that is,  right in the region of interest for the anomalies 
{\bf [A]}---{\bf [C]}.  They are the lowest members of a predicted hidden 
sector which have no direct interaction with our sector as described by the 
standard model.  Of these low mass states $H_{\rm light}$ and $G_{\rm light}$, two are 
singled out, both electrically neutral, one a scalar state $H_+$, and the other 
a vector $G_3$, because they mix, the former with the Higgs boson $h_W$, and 
the latter with the photon $\gamma$, hence acquiring each a small coupling to 
standard sector particles.  The scalar $H_+$ with a small admixture of $h_W$ 
can play the role of $U$, contributing to lepton magnetic dipole moments 
{\bf [A]} via the diagram in Figure \ref{g-2} and to Lamb shifts {\bf [B]} via the 
diagram in Figure \ref{Lambsfig}, and since it couples to leptons like the Higgs 
$h_W$, it favours the muon over the electron, thus helping to explain why these 
two anomalies are observed only for the muon, not for the electron.  The vector 
$G_3$ with a small admixture of $\gamma$, on the other hand, is predicted to have 
a mass of 17 MeV, that is,  right on top of the $X$ required for explaining the 
Atomki anomaly {\bf [C]}. 

Unfortunately, however, FSM predictions at present do not, strictly speaking, 
go much beyond the above observations.  To proceed further depends on some 
parameters of the model the values of which are still unknown, and on some 
loop calculations for which the technology has not yet been developed.  For 
this reason, our present investigation falls short of an actual explanation of 
the anomalies.  What can be aimed for at present is only an examination of 
whether the anomalies can be accommodated in the FSM framework.  In this, the 
answer seems positive, as we shall demonstrate.

Despite this limitation, a positive point should be noted for the  FSM.  In contrast
to many, perhaps even most, of the models so far suggested for addressing these 
anomalies, the FSM was {\em not} created for this purpose.  The FSM was suggested 
in the first place  to explain the generation puzzle, namely why there should be 3 and only
3 generations of quarks and leptons and why they should fall into a hierarchical
mass and mixing pattern.  This it has done, we think, quite well as summarized
in a recent descriptive review \cite{fsmpop}.  Briefly, the FSM assigns to 3 
generations a geometrical significance as the dual to colour and obtains the 
observed mass and mixing patterns of quarks and leptons as a result of a mass
matrix rotating with scale, which is itself a consequence of renormalization 
group equations.  In practical terms, it replaces effectively 17 parameters of 
the standard model by 7, reproducing about a dozen mass and mixing parameters
mostly to within present experimental errors \cite{fsmpop,cfsm,tfsm}.  For
doing so, the FSM has added to the SM some new field variables called framons 
which are similar in concept to the vierbeins in gravity, and it is these 
framons which are manifesting themselves now as new particles in a ``hidden 
sector'', of which the $H_+$ and $G_3$ suggested above for accommodating the 
anomalies are specific examples.  The structure of the model and most of its 
parameters are thus already fixed by the requirement to fit the standard sector.
  What can be adjusted to accommodate the anomalies of interest to us here is 
only the remaining freedom.  It is, in other words, rather a strait-jacket 
to fit.  That it can be done at all seems thus quite nontrivial.  For example, 
that the model puts the $G_3$ or $X$ mass at 17 MeV is a genuine prediction of 
the model based on the fit to quark and lepton data done in \cite{tfsm} which 
predates the discovery of the Atomki anomaly.

Before starting on the anomalies, there is first an important simplification to 
note, arising from the following two facts:
\begin{itemize}
\item That $H_+$ having $J^P = 0^+$  cannot appear as $X$ in (\ref{Be*decay})
if parity and angular momentum are to be conserved, since ${}^8{\rm Be}^*$ is $1^+$
and ${}^8{\rm Be}$ is $0^+$.
\item That the contribution of $G_3$ to the diagrams in Figure
  \ref{g-2} and Figure \ref{Lambsfig} 
which give the anomalies {\bf [A]} and {\bf [B]} is likely to be small compared
to that of $H_+$ because the mixing of $H_+$ to $h_W$ which gives its coupling 
to leptons occurs at tree level while that of $G_3$ to $\gamma$ occurs  at 
1-loop.
\end{itemize}
To a first approximation therefore, only $H_+$ need be considered for {\bf[A]} 
and {\bf [B]}, and only $G_3$ for {\bf [C]}.  We adopt this approximation in 
what follows, but shall be able afterwards to check its consistency.

The result of our analysis has already been summarized in the abstract, which
need not be repeated.  But there is here one extra point worth noting.  The 
FSM as a theory being only at an embryonic stage, not all---in fact not too 
many---of the conclusions drawn from it so far can claim to be logically tight 
or beyond reasonable doubt.  This is true especially in the ``hidden sector'' 
where experimental information is hard to come by, and FSM is much in need of 
phenomenological support.  However, in the following two sections, 2 and 3, 
while trying to accommodate the anomalies in the FSM framework, we shall take
previous FSM results as granted without any question, so as not to disrupt the 
flow of the argument.  Only in Section 4 shall we return to review the steps 
leading to some of the FSM conclusions used and expose possible weaknesses.  
It will be seen then that by accommodating the anomalies in the FSM, each side 
would lend support to the other, and end up, it seems, in strengthening the 
credibility of both.  And, apart from offering a possible explanation for the 
anomalies, it has implications for the FSM as well, in helping, for example, 
to resolve some ambiguities there.

\section{The $g - 2$ and Lamb shift anomalies}

As worked out in, for example \cite{tfsm}, the state $H_+$ of present interest
mixes at tree-level with the standard electroweak theory Higgs $h_W$ and with
another state in the ``hidden sector'' called $H_3$ via the mass-squared 
submatrix:
\begin{equation}
M =
\pmatrix{
4\lambda_{W}\zeta_{W}^{2} & 2\zeta_{W}\zeta_{S}(\nu_{1}-\nu_{2})
\sqrt{\frac{1+2R}{3}} &
2\sqrt{2}\zeta_{W}\zeta_{S}\nu_{1}\sqrt{\frac{1-R}{3}} \cr
\ast & 4(\kappa_{S}+\lambda_{S})\zeta_{S}^{2}\left(\frac{1+2R}{3}\right) & 
4\sqrt{2}\lambda_{S}\zeta_{S}^{2}\frac{\sqrt{(1+2R)(1-R)}}{3} \cr
\ast & \ast &
4(\kappa_{S}+2\lambda_{S})\zeta_{S}^{2}\left(\frac{1-R}{3}
\right) \cr
},
\label{Msubmatrix}
\end{equation}
where the asterisks  $\ast$ denote symmetric entries, 
$\lambda_W, \lambda_S, \kappa_S, \nu_1, \nu_2$ are coupling coefficients
of various terms in the scalar framon self-interaction potential, $\zeta_W$ 
is the vacuum expectation value  of the flavour framon (in other
words,  the electroweak Higgs field), and
$\zeta_S$ the vacuum expectation value of its colour
analogue\footnote{Note 
  that in FSM, as in SM, colour is confined and colour symmetry remains
exact.  We recall that in the confinement picture of ‘t Hooft for the
electroweak theory \cite{thooft}, the local symmetry $su(2)$ is also confining and exact,
and $\zeta_W \neq 0$ means only that an associated (dual) global symmetry,
say $\widetilde{su}(2)$, is broken.  Similarly then in FSM, that the colour
framon has a nonzero vacuum expectation value $\zeta_S$
means only that the global (dual colour)
symmetry, $\widetilde{su}(3)$ representing generations for fermions, is
broken while local colour symmetry remains exact \cite{cfsm}.}.
Of these quantities, which may all
depend on scale, one knows at present only the values of $\lambda_W,\zeta_W$ 
and an estimate for $\zeta_S$ at $\mu \sim m_Z$, and little about the
scale-dependence of any.  

\begin{figure}
  \centering
\includegraphics[scale=0.4]{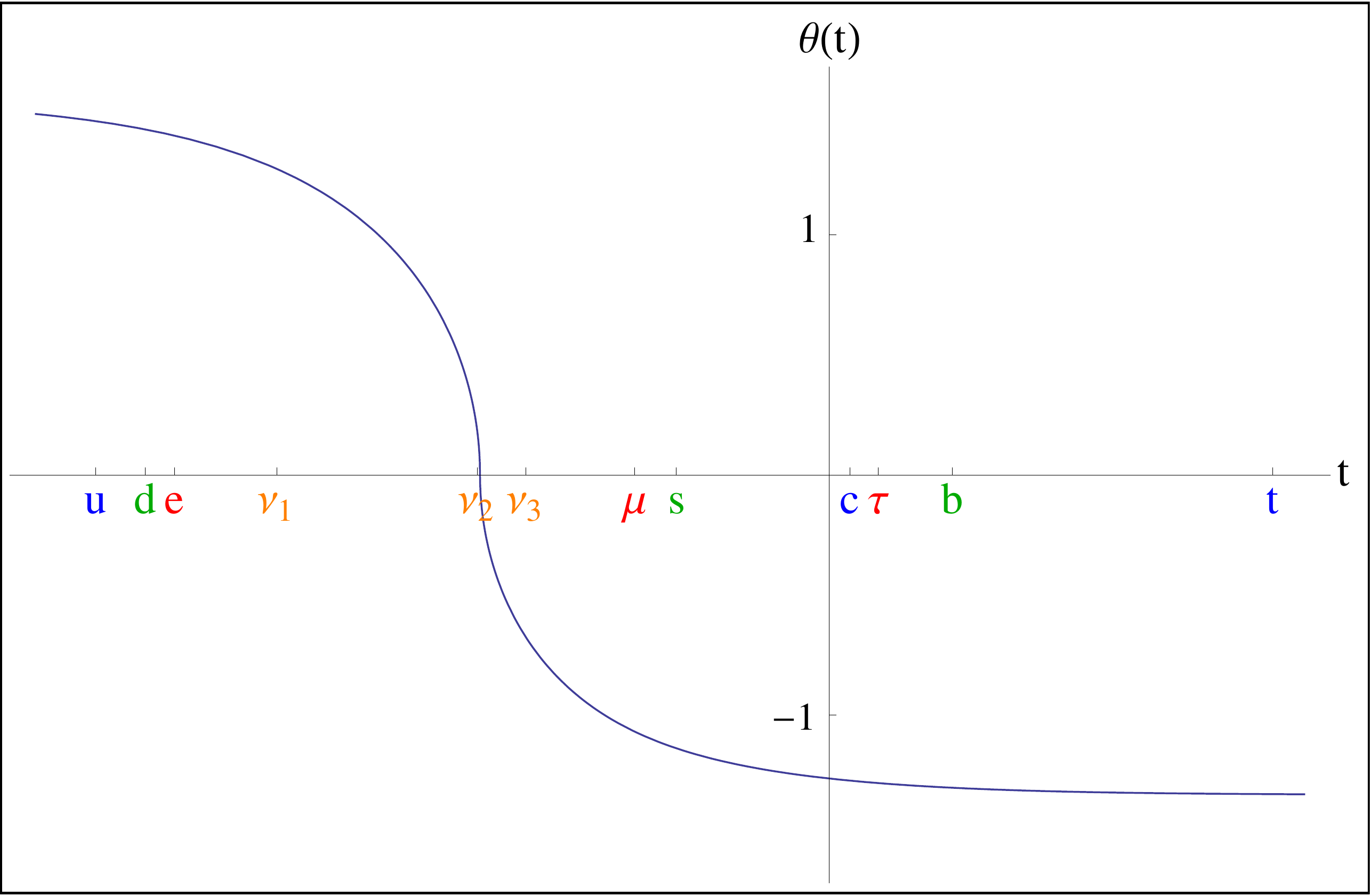}
\caption{The polar angle $\theta$ of the normalized rotating vector $\balpha$ as
  a function of $t=\ln \mu$ ($\mu$ being the scale) obtained in
  \cite{tfsm} by fitting quark and lepton data, where the
  azimuthal angle $\phi$ is given
by $\cos \theta \tan \phi =-0.1$}
\label{balpha}
\end{figure}

\begin{figure}
  \centering
 \includegraphics[scale=0.4]{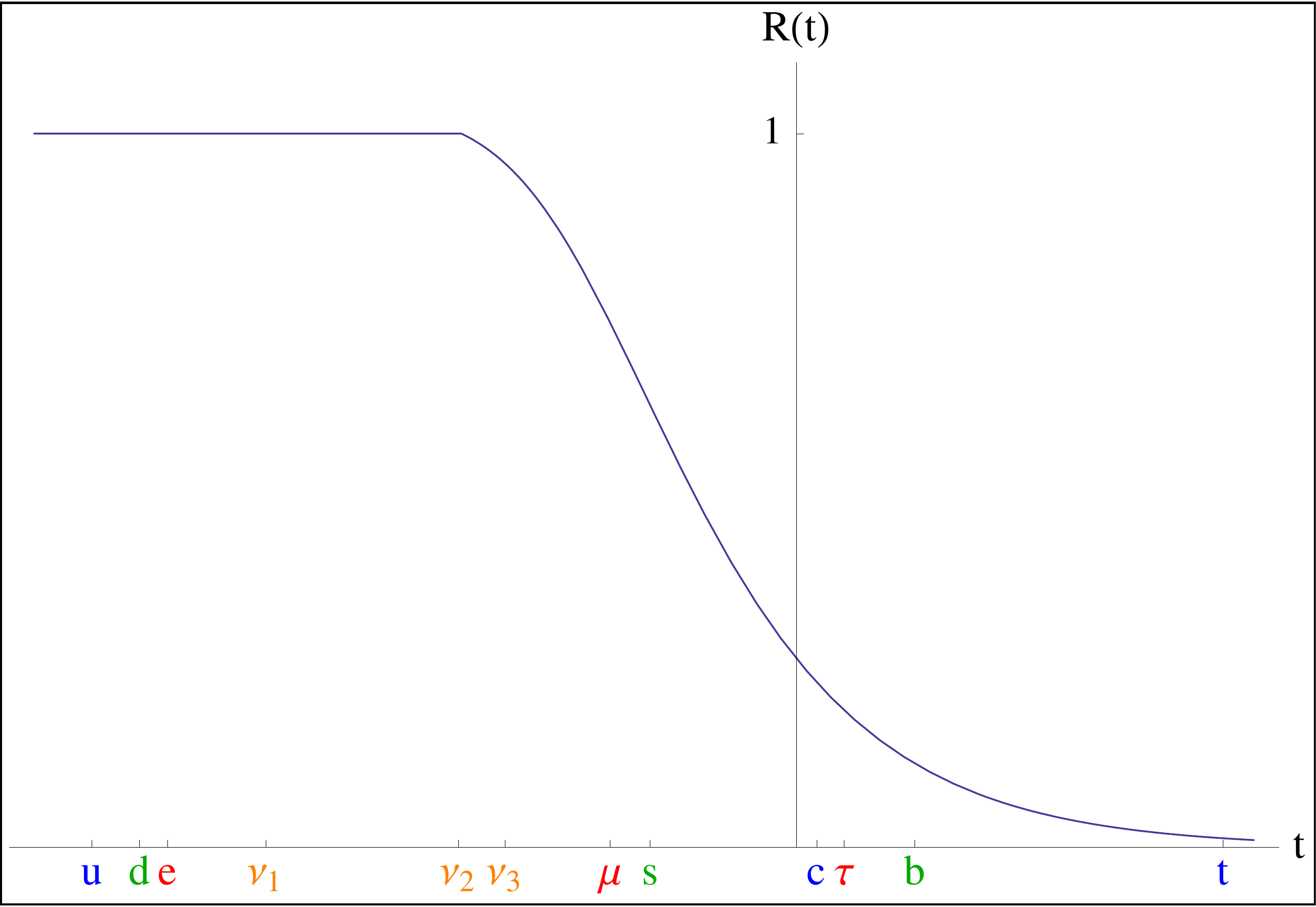} 
\caption{The ratio $R = \nu_2 \zeta_W^2/ 2 \kappa_S \zeta_S^2$ as a function of 
$t=\ln \mu$ ($\mu$ being the scale)
  obtained by the same fit to quark and lepton data in \cite{tfsm}}
\label{R}
\end{figure}

An exception is the combination $R = \nu_2 \zeta_W^2/ 2 \kappa_S \zeta_S^2$ 
which occurs frequently in the FSM.  It is a measure of the relative strength 
of the symmetry-breaking versus the symmetry-restoring terms in the framon 
potential for what is for quarks and leptons the generation symmetry, called
$\widetilde{su}(3)$ in FSM.  Its value and scale-dependence is correlated to
that of the vector $\balpha$, the rotation of which with changing scale is 
what gives the hierarchical mass and mixing patterns of quarks and leptons,
as mentioned in the Introduction.  Thus the fit to these patterns \cite{tfsm}
has already supplied us with the information of how $\balpha$, and hence also
$R$ depend on scale.  This is reproduced in Figures \ref{balpha}, \ref{R} for
easy future reference.

At tree level, were it not for the said mixing, $h_W$ would have the mass 125 GeV 
observed in experiment, $H_3$ would have a mass of order TeV \cite{cfsm,zmixed},
 and $H_+$ a mass of around 17 MeV, the same as its diagonal partners in 
$H_{\rm light}$ and $G_{\rm light}$, such as the $G_3$ or $X$ to be considered in Section 
3 on the Atomki anomaly.  The reason for such an abnormally small mass for
these states is their having mass squared matrix elements proportional to
$1 - R$, as seen for example in (\ref{Msubmatrix}), which, according to Figure
\ref{R}, vanishes sharply at around 17 MeV and forces a 
solution near that value \cite{tfsm} to the condition:
\begin{equation}
m_x(\mu) = \mu
\label{mphys}
\end{equation}
for the physical mass of any state $x$ by normal convention as a fixed point
of the function $m_x(\mu)$.\footnote{There is another solution of (\ref{mphys})
at scales of order TeV, but this would be unstable decaying into the former.} 

When mixing by (\ref{Msubmatrix}) is taken into account, the masses of all 3
states $h_W, H_3, H_+$ will change but are still expected to be close to the 
above values.  In particular, the mixed state $U$ from $H_+$ would have a 
mass $m_U$ close to but above 17 MeV, say of order 20 MeV.

As to exactly how the mixing works out for (\ref{Msubmatrix}), it is not as 
straightforward as it might seem at first sight.  Since the parameters on which
the matrix depends are themselves dependent on scale, some strongly such as 
$R$, and we are dealing with scales differing by some 5 orders in magnitude, 
it is unclear at what scale or scales that the matrix should be diagonalized. 
In our opinion, a stepwise diagonalization procedure, as adopted before for the 
rotating mass matrix in say \cite{tfsm}, would seem appropriate.  But we 
are unsure, not having in the ``hidden sector'' the same abundance of 
data to check against as in the earlier case in the standard sector.  Besides, 
even if we just  proceed as before, the answer will not be informative 
because the values of most parameters in (\ref{Msubmatrix}) are still unknown 
or known only at certain scales.

Fortunately, for the problem at hand, little needs be known about the details
of the mixing, so long as it is accepted that $U$ will acquire a component in
$h_W$, say $\rho_{Uh} h_W$, where $\rho_{Uh}$ is presumably small which we can 
treat as an adjustable parameter to be fitted to experiment.  We recall that 
$H_+$ has itself no direct couplings to quarks and leptons, so that $U$ will 
couple to quarks and leptons only via its $h_W$-component, and have thus the 
same couplings to these as $h_W$ does, only at a different scale and with 
strength reduced by the factor $\rho_{Uh}$.

The standard Higgs state $h_W$, in FSM as in SM, couples to quarks and leptons 
via the Yukawa term, but this term differs in details in the two models.  In the
SM, there is introduced a Yukawa term for each quark or lepton with a coupling
proportional to the fermion mass.  In the FSM, there is introduced only
one Yukawa term for all three generations of each quark or lepton species, 
thus:
\begin{equation}
\sum_{[\tilde{a}],[\tilde{b}]} Y_{[\tilde{b}]} \bar{\psi}^{[\tilde{a}]}_L  \alpha_{[\tilde{a}]}
   \phi \psi^{[\tilde{[b}]}_R,
\label{Yukawa}
\end{equation}
where the vector $\balpha$ is an attribute of the flavour framon which 
replaces the Higgs scalar field in the standard electroweak theory, and hence 
is independent of the species of quarks or leptons to which (\ref{Yukawa})
refers.

Expanding (\ref{Yukawa}) in fluctuations of the $h_W$ field about its
vacuum expectation value $\zeta_W$, we obtain to zero-th order the mass matrix
depending on the rotating vector\ \balpha, which gives in
FSM the mass and mixing patterns of quarks 
and leptons mentioned in the Introduction \footnote{See {\bf [R3]} of Section 
4 for a brief explanation, and earlier papers for more details.}.
Then to first order,
\begin{equation}
\mathcal{L}_Y \, = \, - 
\sum_{T=U,D,L,\nu} \frac{m_T}{\zeta_W} \sum_{i=1,2,3} 
|\langle \balpha(\mu) |f_i \rangle|^2 \, \, \overline{f_i}{f_i} \, h_W.
\label{hcouplings}
\end{equation}
In (\ref{hcouplings}) then, only the coefficient $m_T$ depends on the
species, this being the mass of the heaviest generation of that species, thus
for the $U$-type quarks, $m_T = m_t$, the mass of the top quark, and for the 
charged leptons $m_T = m_\tau$, the mass of the tau lepton.

According to previous arguments then, the couplings of $U$ to quarks and 
leptons differ from (\ref{hcouplings}) only by being multiplied by the mixing 
factor $\rho_{Uh}$ and by the fact that for the problems at hand the rotating 
scale-dependent vector $\balpha$ is to be evaluated, by usual prescription, 
at the scale of the $U$ mass, $m_U$.  

The fit \cite{tfsm} to mass and mixing data for quarks and leptons has already 
supplied us with the information how $\balpha$ depends on scale $\mu$ as shown 
in Figure \ref{balpha} from which $\balpha(m_U)$ can be read off.  The same fit 
has also given us the state vectors $|f_i \rangle$ for all the quarks and 
leptons; in particular those for $\mu$, $e$ and $u$ which we shall need 
here are reproduced below for easy reference:
\begin{eqnarray} 
\langle \mu| & = & (0.07434, -0.53580, 0.84107), \nonumber \\  
\langle e| & = & (0.42028, 0.78168, 0.46082), \nonumber \\ 
\langle u| & = & (0.42041, 0.53974, 0.72934).
\label{langlemueu} 
\end{eqnarray} 

\subsection{$g - 2$}

\begin{figure}
  \centering
  \includegraphics[scale=0.5]{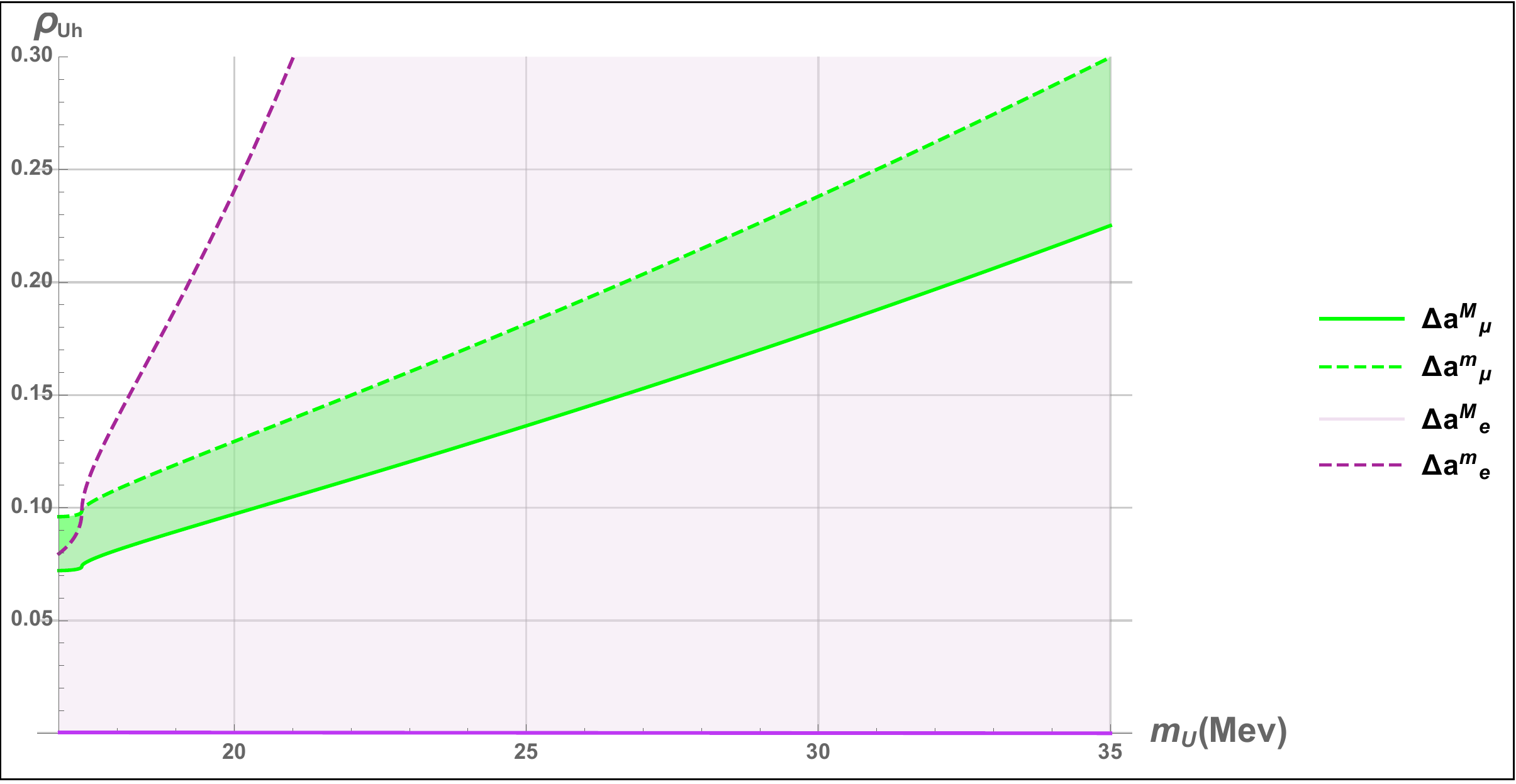}
\caption{Experimental  bounds on the mixing parameters  $\rho_{Uh}$
  from $g-2$ for muon (green) and electron (mauve), where the allowed
  regions are shaded} 
\label{rhoUhmue} 
\end{figure}

The contribution of Figure \ref{g-2} to the $g - 2$ anomaly $\Delta a_\ell$ is 
of the form \cite{integral}:
\begin{equation}
\Delta a_{l} \, = \, \frac{g_{U\ell}^2}{8 \pi^2} \,  \int_0^1 dz \, 
\frac{(1+z)(1-z)^2}{(1-z)^2 + z \left( \frac{m_{U}}{m_\ell} \right)^2}.
\label{Deltaal}
\end{equation}
Given then the experimental bounds (\ref{g-2}) on $\Delta a_\mu$, one obtains 
the corresponding bounds on $g_{U \mu}$ for any value we choose for $m_U$. We 
choose instead to display these bounds, using:
\begin{equation}
g_{U\ell} = \rho_{Uh} \frac{m_\tau}{\zeta_W} |\langle \balpha(\mu)|\ell \rangle|^2
\label{gUl}
\end{equation}
from (\ref{hcouplings}), in terms of $\rho_{U h}$, for easier comparison 
later with other channels, and obtain the result in Figure \ref{rhoUhmue}.
One sees that in the expected range of $m_U$ around 20 MeV, $\rho_{Uh}$ is found 
to have value of order $10^{-1}$.

This conclusion can actually be deduced qualitatively without any explicit 
calculation as follows.  For $m_U \sim 23$ MeV, and $m_\mu = 105$ MeV, the 
$m_U$-dependent term in the integral in (\ref{Deltaal}) is negligible so that 
the integral is of order unity and approximately constant.  The quantity 
$|\langle \mu|\balpha \rangle|^2 m_\tau/\zeta_W$ in (\ref{gUl}) has a
value \footnote{This estimate depends, via
  $\balpha$, on the  value taken for
$m_U$.  For this and all subsequent estimates of $U$ couplings in this
section, we have taken as benchmark $m_U = 23$ MeV which, as will be seen in
subsection 2.3, is the smallest value allowed by present experimental bounds.   For
$\mu$ at this value, $\balpha^\dagger = (-0.47971, 0.05478, 0.87571)$ in the
fit of \cite{tfsm} which gives Figure 3.}  of 
order $3 \times 10^{-3}$ ,
which gives immediately 
\begin{equation}
\rho_{Uh}^2 \sim 8 \pi^2 \times 10^5 \times 10^{-9} \sim 10^{-2} 
\label{FUhest} 
\end{equation} 
as claimed.

One notes here then a point:
\begin{itemize}
\item {\bf [n1]} the value needed for $\rho_{Uh}$ is of order $10^{-1}$,
(that is,  $< 1$ and small), which is consistent with its interpretation in FSM
as a mixing parameter. 
\end{itemize}

The experimental bound on $\Delta a_e$, we recall from (\ref{g-2}), is some
3 orders lower than that for $\Delta a_\mu$.  Most of this suppression is due 
just to kinematics, as seen in the integral (\ref{Deltaal}), where now for 
$m_U \sim 23$ MeV and $m_e \sim 0.5$ MeV, and thus $m_U/m_e \sim 46$, the 
$m_U$-dependent term completely dominates, giving the integral approximately as 
$\sim (m_e/m_U)^2 \ln(m_e/m_U)$, that is,  already some 3 orders of magnitude lower. 
However, this is still not quite enough to satisfy the experimental bound which 
needs $g_{U\mu} > g_{Ue}$ by
about another factor of around 3---6.  And this is aptly supplied by the 
difference between $\langle e|\balpha \rangle$ and $\langle \mu|\balpha \rangle$
 in (\ref{langlemueu}) above.  That this is so can of course be confirmed by 
detailed calculation as shown in Figure \ref{rhoUhmue}.

One notes then another point:
\begin{itemize}
\item {\bf [n2]} For any value of $\rho_{Uh}$ allowed by experiment on the 
muon $g - 2$ anomaly in the range of $m_U$ under consideration, the FSM 
predicts a value for the electron $g - 2$ anomaly which automatically satisfies 
the bound set by experiment.  This tests the FSM prescription in
(\ref{gUl}) for $U$ (and eventually
(\ref{Yukawa}) for $h_W$) for the relative sizes of the couplings to different 
generations of the same species (in this case the charged leptons).
\end{itemize}

\subsection{Lamb Shift}

The contribution of $U$ via Figure \ref{Lambsfig} to the splitting of the 2S-2P 
energy levels of leptonic atoms is given as \cite{lambshiftformula}:
\begin{equation}
\delta E_\ell^A \, = \, -  \,g_{U\ell} \, g_{UN} \, 
\frac{m_{U}^2 a_{0N}}{8 \pi\left(1+m_{U} a_{0N}\right)^4}
\label{deltaElN}
\end{equation}
where $g_{UN}$ is the coupling of $U$ to the nucleus $N$, with mass
$m_N$ and charge $Z$, of atom $A$
for which the Lamb shift
is being considered, and
\begin{equation}
a_{0N} \, = \, \frac{1}{Z \alpha} \left( \frac{1}{m_N} + \frac{1}{m_\ell} \right)
\label{Bohrrad}
\end{equation}
is the Bohr radius, where $\alpha$ is
the fine structure constant.

The formula (\ref{hcouplings}) gives the $U$ couplings to quarks as:
\begin{equation}
g_{Uq} = \rho_{Uh} \frac{m_{t,b}}{\zeta_W} |\langle \balpha(m_U)|q \rangle|^2
\label{gUq}
\end{equation}
which are similar in form to those to leptons in (\ref{gUl}) except for the 
replacement of the factor $m_\tau$  by $m_t$ or $m_b$, depending on whether
we are dealing with up-type or down-type quarks.  To calculate $\delta E_\ell^A$ 
in (\ref{deltaElN}), however, we need instead the $U$ couplings to the proton 
$p$ or possibly other nuclei.  These we may write as:
\begin{eqnarray}
 g_{UN} & = & \rho_{Uh}
\frac{m_t}{\zeta_W} \, \sum_{q = u,c,t} \langle N \mid  \overline{q} q
              \, \mid N \rangle  
   |\langle \balpha(m_U)|q \rangle|^2 \nonumber \\ 
       & + & \rho_{Uh}
\frac{m_b}{\zeta_W} \, \sum_{q= d,s,b} \langle N \mid  \overline{q} q
             \, \mid N \rangle
   |\langle \balpha(m_U)|q \rangle|^2.
\label{gUN}
\end{eqnarray}
Given that $m_t \gg m_b$, we can drop the second term which is some 2 orders 
down in magnitude.  Furthermore, given that the nucleon or deuteron contains 
rather little $c\bar{c}, t\bar{t}$, we can to a good approximation
keep only one term in the 
sum of (\ref{gUN}), and write:
\begin{equation}
 g_{UN}  \, = \, \rho_{Uh}  \frac{m_t}{\zeta_W}
\mid \langle \balpha(m_U) \mid u \rangle\mid^2 \langle N | \overline{u} \, u \, | N \rangle.
\label{gUN1}
\end{equation}
That one can keep only the $u$ term in (\ref{gUN}) is a great simplification
specific to the FSM which will be seen to give explicit physical consequences.

To evaluate this, one needs the matrix element $\langle N |\overline{u} \, u \,| N \rangle$
involving nonperturbative physics, for which one has at present to rely on 
lattice calculations or on phenomenological methods.  In Table \ref{lattice+}
we list a sample selection we found in the literature giving rather different
values for the matrix elements, reflecting presumably the uncertainty in the 
calculation.

\begin{table}
\centering
\begin{tabular}{c||c|c|c}
Ref. & $\langle p|\bar{u} u|p \rangle$ & $\langle n|\bar{u} u|n \rangle$ & 
       $\langle d|\bar{u} u|d \rangle$ \\
\hline
(a)  & 2.10 & 1.31 & 3.38 \\
(b)  & 2.8  & 2.2  &  ---    \\
(c)  & 5.2  & 4.3  &  ---    \\
(d)  & 6.8  & 6.0  &  ---    \\
\hline
\end{tabular}
\caption{Relevant matrix elements at 2 GeV  taken from lattice calculations
  (a) \cite{Chang}, (b)  \cite{Abdel-Rehim},
(c) \cite{Alexandrou}, and (d) phenomenological estimates
\cite{Grinstein}}
\label{lattice+}
\end{table}

With the matrix elements $\langle N|\bar{u}u|N \rangle$ taken from
Table \ref{lattice+},
divided by the factor 1.33 to take account of the running from 2 GeV to
the
nucleon mass scale at 1 GeV \cite{running}, one obtains then an estimate
for the
coupling $g_{UN}$.  We shall take Reference (a) as benchmark, this
being the
only one to give an estimate for the deuteron and incidentally also the
most
recent, but the parallel results for cases (b), (c), and (d) can be
readily
inferred from those of (a).

Before we do so, however, we note first the following.  One recalls
from (\ref{Lambshift})
that the value of $\delta E_\mu^H$ measured in experiment is about $-0.3$
meV.
With all other quantities in the expression (\ref{deltaElN})  already known, one can
then
estimate what sort of coupling $g_{Up}$ is needed to reproduce such a
value
for $\delta E_\mu^H$.  The answer is that, rather surprisingly at first
sight,
$g_{Up}$ must be some 2 orders of magnitude larger than $g_{U \mu}$.
That this
is so can be seen as follows.  The quantity $m_U a_{0N}$ in
(\ref{deltaElN})  for
$m_U \sim$
23 MeV has a value of about 33, so that:
\begin{equation}
 \delta E_\mu^H \sim - g_{U \mu} g_{U p} (8 \pi)^{-1} (33)^{-3} m_U,
\label{deltaEmup}
\end{equation}
which works out as $g_{Up} \sim 3.0 \times 10^{-2}$, a factor some 2
orders
larger than $g_{U \mu}$ which is given by (\ref{gUN1}) as $4 \times 10^{-4}$,
as was
claimed.  This conclusion is obtained only from (17) and the data.

On the other hand, applying instead the formula (21) from FSM by the
procedure
outlined in the paragraph before last, one obtains the estimate $g_{Up}
\sim
2.95 \times 10^{-2}$, magically coinciding with the estimate above from
data.
The 2 orders of magnitude difference is aptly reproduced by the FSM in
that the
coupling $g_{Up}$ in (21) contains a factor $m_t$, in place of $m_\tau$
of the
coupling $g_{U \mu}$ in (\ref{gUl}), and indeed $m_t/m_\tau \sim 10^2$.

We note thus another point:
\begin{itemize}
\item {\bf [n3]} To accommodate the experimental value for $\delta E_\mu^H$ in 
(\ref{Lambshift}), the coupling $g_{Up}$ of $U$ to the proton has to be some two 
orders of magnitude greater than that to the muon $g_{U\mu}$, which is aptly 
reproduced by the FSM, because of $u$-dominance (\ref{gUN1}) and
the fact that $m_t\gg m_\tau$.
This checks near quantitatively the FSM prescription 
(\ref{hcouplings}) for the $U$ (or $h_W$) couplings to fermions of different species (in 
this case the up-type quarks versus the charged leptons).
\end{itemize}

Next, we turn to the Lamb shift in muonic deuterium given in (\ref{Lambshift}),
where it is seen that $\delta E_\mu^D$ is less than twice that for muonic 
hydrogen $\delta E_\mu^H$, implying by (\ref{deltaElN}) that 
\begin{itemize}
\item {\bf [n4]} the coupling $g_{U n}$ of $U$ to the neutron is  less than
that to the proton $g_{Up}$, and that this immediately follows from the FSM
prediction (\ref{gUN1}) of dominance by the $u$ quark and that the neutron has
less $u$ quarks than the proton.  This checks again the FSM prescription 
(\ref{hcouplings}) for $U$ (or $h_W$) couplings to different fermion species (this time 
up-type quarks versus down-type quarks).
\end{itemize}

The same analysis can be repeated for the electron to obtain the Lamb shift
for ordinary hydrogen and deuterium.  Again, given that the Bohr radii are
here some 200 times larger than for muonic hydrogen and deuterium, the Lamb
shifts are also appropriately suppressed, and these effects, together with 
the smaller $U$ couplings to the electron than to the muon, are enough to
bring the anomaly below present experimental sensitivity.  Hence

\begin{itemize}
\item {\bf [n2']} Given $\rho_{U\ell}$ in Figure \ref{rhoUhmue}, the FSM predicts
values of the Lamb shift anomalies for the electron which are below present
experimental detection sensitivity, as is indeed the case.
\end{itemize}

\subsection{$g - 2$ and Lamb shift combined}

\begin{figure}
  \centering
  \includegraphics[scale=0.5]{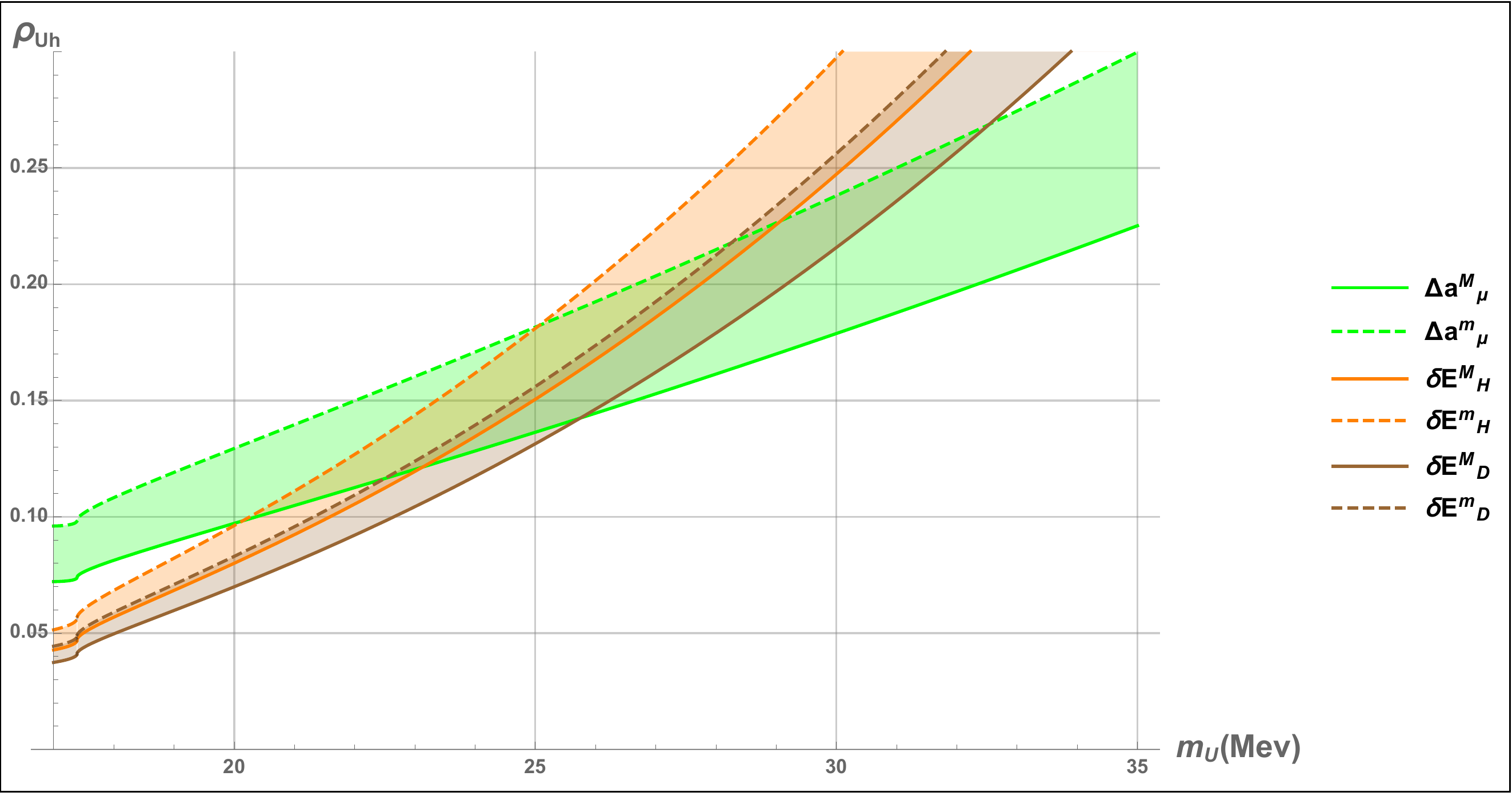}
\caption{Experimental  bounds on the mixing parameters  $\rho_{Uh}$
  from muonic $g-2$ (green), Lamb shift in muonic hydrogen (orange),
  and Lamb shift in muonic deterium (brown), where the allowed regions
are shaded}
\label{Lambshiffc}
\end{figure}

Inputting next the experimental bounds in (\ref{Lambshift}) on $\delta E_\mu^H$,
and using (\ref{deltaElN}) and the formulae (\ref{gUl}) and (\ref{gUN1}), one
obtains, for (a) in Table \ref{lattice+}, the bounds on $\rho_{Uh}$ displayed 
in Figure \ref{Lambshiffc}.  The same is done with the Lamb shift in deuterium
and shown together with the earlier bounds obtained from the $g - 2$ anomaly.  
One notes that all three bands overlap in the range 23 - 29 MeV, which means
that 
\begin{itemize}
\item {\bf [n5]}  For all bounds on muonic $g - 2$ and Lamb shift to be 
satisfied, $m_U$ has to be in the range 23 - 29 MeV, right in the region 
expected by (\ref{mphys}).
\end{itemize} 
Had one chosen to work with (b), (c) or (d) of Table \ref{lattice+} instead of 
(a), it would mean just parallely shifting the Lamb shift bands a little to the 
right to higher values in $m_U$, but still broadly in the expected range.

It is interesting that in coming to the conclusion of a mass for $U$ of order
20 MeV in {\bf [n5]}, one has made no use of the mass condition (\ref{mphys})
at all.  The conclusion was drawn merely from the data on the two anomalies 
and the assumption that they both arise by  virtue of the same $U$ boson.
Then  the requirement of consistency between the two sets of data already leads 
to the result $m_U \sim 20$ MeV.  One may thus regard {\bf [n5]} as an 
independent check by data on a qualitative prediction of the FSM.

Conversely, given that $m_U \sim 20$ MeV is already expected from 
(\ref{Msubmatrix}) and (\ref{mphys}),
which assertion can in principle be confirmed in future when 
enough information on the parameters involved becomes available to allow the 
solution of (\ref{mphys}), then the same result can be rephrased in summary of 
this section as: 
\begin{itemize}
\item {\bf [n6]}  By adjusting a single free parameter $\rho_{Uh}$, the FSM has 
accommodated the following 6 independent pieces of experimental information: 
$$\Delta a_\mu, \Delta a_e, \delta E_\mu^H, \delta E_\mu^D, \delta E_e^H, 
\delta E_e^D,$$ on two distinct physical effects, namely $g - 2$ and Lamb shift, 
which need, {\it a priori}, have nothing to do with each other.
\end{itemize}

However, the above analysis, seemingly successful so far, has difficulty in 
explaining the apparent absence of a Lamb shift anomaly in some preliminary 
data on muonic helium \cite{Heexpt,nucleonbinding}.
Also it gives mixed results  when 
applied to  bounds involving heavy nuclei:
\begin{itemize}
\item Nucleon binding energy in nuclear matter.
\begin{equation}
\Delta B = 
\frac{1}{4 m_U^2} \left( g_{Up} \, + \,  g_{Un} \right)^2 
\rho
\label{deltaB}
\end{equation}
$\rho=0.64 \times 10^6$ MeV$^3$. The experimental limit
\cite{nucleonbinding}
is $\Delta B < 1$  MeV.
\item  In neutron scattering on lead, the quantity to consider is
(assuming
coherence in the nuclear-scalar interaction) the effective coupling
\cite{neutronleadthe}:
\begin{equation}
g_{Un} \left( \frac{A-Z}{A}g_{Un} \, + \, \frac{Z}{A} g_{Up} \right),
\nonumber
\end{equation}
where A = 208; Z = 82 for Pb.  The experimental limit for this
\cite{neutronleadexp,neutronleadexp2} 
works out to be $5.6 \times 10^{-10} \left(
\frac{m_U}{\mathrm{MeV}}
\right)^4$, giving the bounds, for $m_U=$ 23 MeV:
\begin{equation}
g_{Up} \, < \, 0.018, \quad  g_{Un} \, < \, 0.011,
\label{nPbbounds}
\end{equation}
\end{itemize}

The values obtained earlier in FSM, namely: $g_{Up} = 0.030, g_{Un} =
0.018$,
pass the former bound giving $\Delta B \sim 0.6$, but exceed the latter
bound by
a factor of about 1.6.
These are traditional difficulties for attempts to explain the Lamb shift
anomaly by a new particle as the present case.  To us, it points to some
nonlinear effects which spoil the simple picture adopted above of the lepton
interacting individually with the quark constituents.  This might work to a
certain extent in a nucleon, although, even there, one is troubled by 
uncertainties as indicated by the very different results listed in Table 
\ref{lattice+}.  But for a complex nucleus, such an approximation might become
untenable.  Indeed, even for the deuteron, one already begins to see some 
strain, noting the small overlap in Figure \ref{Lambshiffc} between the hydrogen
and deuterium bands, which, if the approximation works perfectly, should be 
identical.  This seems to us to be a problem in
nuclear physics
to which we have at present no answer.
Later on, in Section 3, we shall encounter another possible
complication due to
compound couplings of the $U$ to nucleons, which might interfere with
the above
couplings $g_{U_{p,n}}$ and further affect these bounds.

\section{The Atomki anomaly}

As already noted in the Introduction, the $U$ boson considered in the preceding 
section being scalar with $J^P = 0^+$, cannot function as the anomaly $X$ in 
Atomki if parity and angular momentum are conserved.  We switch our attention 
therefore to the vector state called $G_3$ in \cite{cfsm}.  This $G_3$ is 
already diagonal at tree-level with mass squared eigenvalue $m_3^2=(g_3^2/6) 
(1 -R)\zeta_S^2$, where $g_3$ is the colour gauge coupling, $\zeta_S$ the 
vacuum expectation value  of the colour framon, and $R$ the scale-dependent 
ratio of couplings displayed in Figure \ref{R}.  As for the $H_+$ or $U$ boson, 
the physical mass of $G_3$ is to be given by solving the fixed point equation 
(\ref{mphys}).  Recalling in Figure \ref{R} that $1 - R$ goes sharply to zero 
near $\mu =$ 17 MeV, and that $\zeta_S$ is bounded in \cite{cfsm} to have a 
value more than 2 TeV at scales of order $\mu \sim m_Z$, it follows that 
(\ref{mphys}) for $G_3$ must have a solution very near to 17 MeV.  There is 
here no complication due to mixing as there is in the parallel problem for 
$H_+$ above.

\begin{itemize}
\item {\bf [n7]}  In other words, the FSM at tree-level predicts quite precisely
 and with little ambiguity a physical mass for $G_3$ right on top of the anomaly
 at 17 MeV observed by Atomki.
\end{itemize}

\begin{figure}
  \centering
  \includegraphics[scale=0.25]{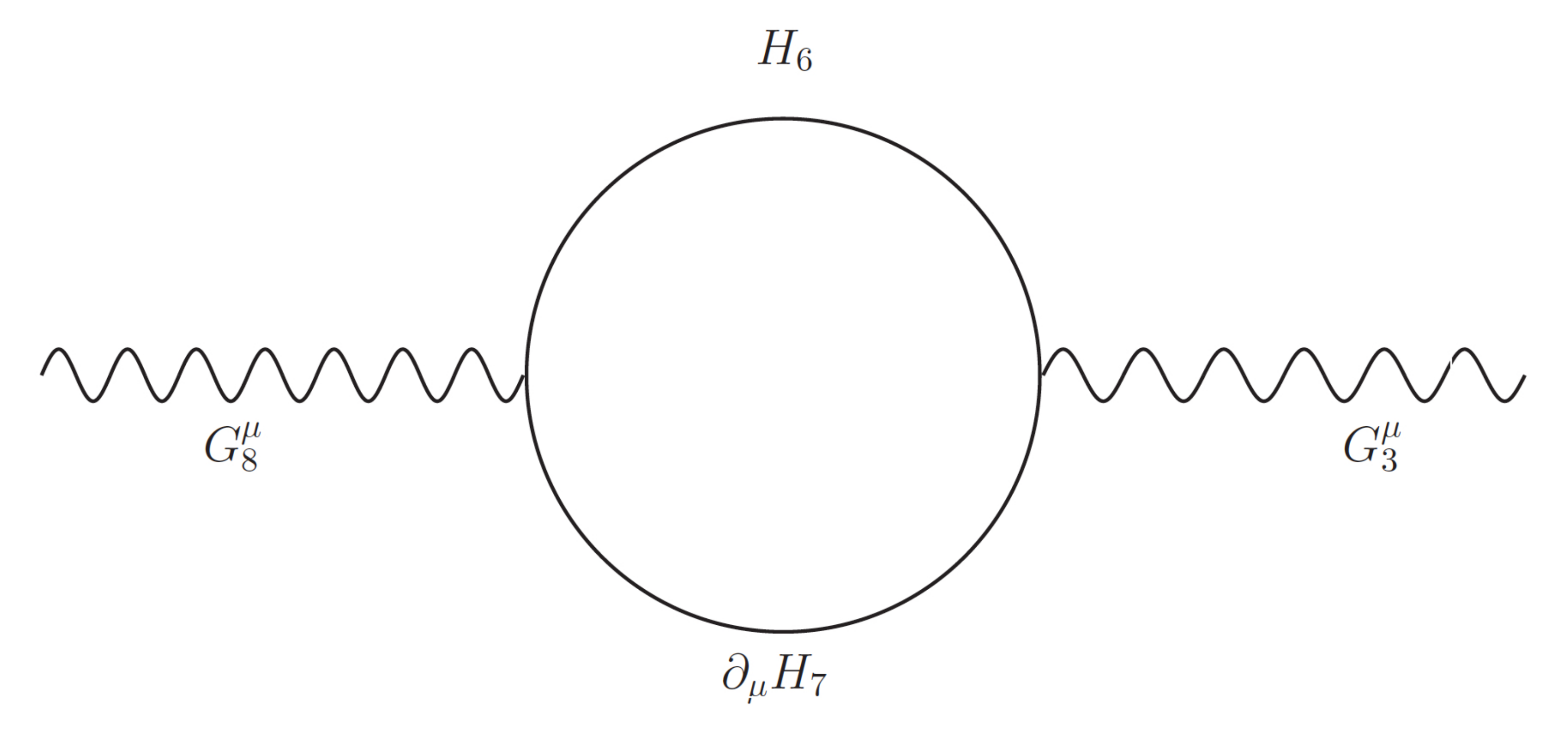}
\caption{Sample 1-loop diagram mixing $G_3$ with $G_8$}
\label{G3G8loop}
\end{figure}

Like the $H_+$, this $G_3$ belongs to the hidden sector in the FSM with no 
direct coupling to standard model particles, and acquires such couplings only
via mixing.  But in contrast to the $H_+$ which mixes with the standard sector
already at tree level, $G_3$ mixes only at 1-loop level.  According to 
\cite{cfsm} where the couplings of one $G$ to two $H$ are listed in Appendix B, 
both $G_3$ and $G_8$ are coupled to $H_6, H_7$ (and also $H_8, H_9$)
so that $G_3$ can 
mix with $G_8$ via the loop displayed in Figure \ref{G3G8loop}.  There are no
doubt similar mixing terms with loops of other particles.  And since it has 
already been shown in \cite{cfsm,zmixed} that $G_8$ mixes with the photon $\gamma$ 
and with $Z$, it follows that this new mixing will induce a mixing of $G_3$ 
with $\gamma$ and $Z$ as well, acquiring for the mixed state $X$ some small 
couplings to quarks and leptons via $\gamma$ and $Z$, a fact that we wish 
now to demonstrate.

In FSM, however, one is not yet in a position to evaluate loop diagrams in 
general, the technology for which has not yet been developed.  One difficulty,
as in the mixing of $U$ in the preceding section, is not knowing at
what scale to do the calculation 
when faced with quantities 
which are strongly scale-dependent.  One cannot therefore perform at present 
a fully consistent calculation to one loop.  Our aim is only to indicate how 
the loop insertion in Figure \ref{G3G8loop} will lead to the mixing of $G_3$
and acquire for the mixed state $X$ a component in $\gamma$.\footnote{Such a 
crudely truncated calculation may not of course preserve gauge invariance nor 
the masslessness of the photon that FSM possesses \cite{zmixed} as a full 
perturbative treatment to 1-loop will, but this will not matter for
the mixing problem in hand.}
And we shall take the loop contribution, which we cannot 
yet calculate, as a small adjustable parameter $\epsilon$ to be treated perturbatively. 

The $4 \times 4$ mass squared matrix for the whole $\gamma-Z-G_8-G_3$ complex 
then appears as:
\begin{equation}
H = H_0 + H',
\label{H}
\end{equation}
with rows and columns labelled by $A_\mu, \tilde{B}_\mu^3, \tilde{C}_\mu^8, 
\tilde{C}_\mu^3$ representing the relevant components of the various fields, 
where
\begin{equation}
H_0 = \left( \begin{array}{cc} M & 0 \\
                               0 & m_3^2 \end{array} \right), 
\label{H0} 
\end{equation} 
and
\begin{equation} 
H' = \left( \begin{array}{cccc} 0 & 0 & 0 & 0 \\
                                0 & 0 & 0 & 0 \\
                                0 & 0 & 0 & \epsilon \\
                                0 & 0 & \epsilon & 0 \end{array} \right), 
\label{Hp} 
\end{equation}

The top left $3 \times 3$ block $M$ has already been diagonalized in 
\cite{zmixed,cfsm}, giving $\gamma, Z, G$ as the eigenstates, corresponding to 
eigenvalues denoted $m_\gamma^2(=0), m_Z^2, m_G^2$.  First order perturbation 
theory for the perturbed $G_3$ state, now identified with $X$ in Atomki, then 
appears as:
\begin{equation}
\psi_{X} = u_{G_3} + \sum_{k = \gamma, Z, G} \frac {\langle k|H'|G_3 \rangle u_k}
  {m_k^2 - m_3^2}.
\label{psiX}
\end{equation}
In the equation above, $u_j$ is the unperturbed state vector of the $jth$ 
eigenstate in the original gauge
basis $A_\mu, \tilde{B}_\mu^3, \tilde{C}_\mu^8, \tilde{C}_\mu^3$.  So $u_{G_3}$ 
is just $(0, 0, 0, 1)$ which has no component in $A_\mu$ and $\tilde{B}_\mu^3$ and 
therefore no coupling to leptons and quarks.  In the second term which 
represents the perturbation, the sum is completely dominated by the  
$k = \gamma$ term because of the denominator, since $m_3^2 \sim (17\
{\rm MeV})^2, m_Z^2 \sim (100 {\rm GeV})^2, m_G^2 \sim ({\rm TeV})^2$.
 Hence, 
\begin{equation} 
\psi_{X} \sim u_{G3} - \frac{1}{m_3^2} \langle\gamma|H'|G_3\rangle u_\gamma.
\label{psiXa}
\end{equation}
Now according to equation (28) of \cite{zmixed}, the state vector $u_\gamma$ for 
$\gamma$ is 
\begin{equation} 
u_\gamma^\dagger = \left(\frac{e}{g_1}, \frac{e}{g_2}, 
                                \frac{2}{\sqrt{3}} \frac{e}{g_3}, 0\right).
\label{ugamma}
\end{equation}
Hence,
\begin{eqnarray}
(\gamma|H'|G_3) & = & u_\gamma^\dagger
                      \left( \begin{array}{cccc}
                                0 & 0 & 0 & 0 \\
                                0 & 0 & 0 & 0 \\
                                0 & 0 & 0 & \epsilon \\
                                0 & 0 & \epsilon & 0 \end{array} \right)
                      \left( \begin{array}{c} 0 \\ 0 \\ 0 \\ 1
                             \end{array} \right)  \nonumber \\
                & = & u_\gamma^\dagger
                      \left( \begin{array}{c} 0 \\ 0 \\ \epsilon \\ 0
                             \end{array} \right) \nonumber \\
                & = & \frac{2e}{\sqrt{3} g_3} \epsilon.
\label{gammaH'G3}
\end{eqnarray}
This shows that the state $\psi_X$ in (\ref{psiXa}) has indeed acquired a
component in $\gamma$ as anticipated.

Finally the factor $u_\gamma$ at the end of (\ref{psiXa}) ensures that it 
couples to quarks and leptons like the photon, except that the coupling is 
reduced by the mixing parameter factor $\rho_{X \gamma} = \frac{2e}{\sqrt{3} g_3} 
\frac{\epsilon}{m_3^2}$.\footnote{It couples like the photon to quarks and 
leptons in FSM which differs a little from the SM but the difference stays 
within present experimental bounds, as shown in \cite{zmixed}.}  In this 
respect, therefore, it behaves as what is often called in the literature a 
dark photon, and as far as the leptons are concerned, it seems that is all we 
need at present.  Thus, it follows for example that $X$ can decay into 
$e^+ e^-$ pairs as is seen in Atomki.

However, $X$ is not just a dark photon, since it is mainly $G_3$,  and 
$G_3$ is not a fundamental field 
in the FSM but a bound state of a colour framon-antiframon pair in $p$-wave 
bound by colour confinement, as interpreted following the confinement picture 
of  't~Hooft \cite{thooft}.  To know how it interacts with the nucleons in the 
beryllium nucleus, which are also composite states bound by colour confinement, 
seems thus a rather complicated affair, probably involving nonperturbative 
effects, a problem one is not in a position to deal with at present, nor 
perhaps for some time to come.  We propose therefore to treat the coupling of 
$G_3$ or $X$ to nucleons as a parameter, say $\kappa_p$ or $\kappa_n$, which 
would be almost the same except that the dark photon-like couplings deduced 
before for charged quarks would still be there and give an extra bit to the 
proton though not to the neutron.  But the difference appears to be small 
compared to the parameter itself, as we shall see later.

In that case, the analysis given in \cite{Feng} applies and gives an estimate 
for the ratio of interest:
\begin{eqnarray}
\frac{\Gamma({}^8 \Be^* \rightarrow {}^8 \Be + (X \rightarrow e^+ e^-))}
  {\Gamma({}^8 \Be^* \rightarrow {}^8 \Be + \gamma)} &\sim&
      \frac{(\kappa_p + 
     \kappa_n)^2}{e^2} S_K \br(X \rightarrow e^+ e^-) \nonumber \\
     &\sim& 5.8 \times 10^{-6}, 
\label{kappa} 
\end{eqnarray} 
where $S_K$ is a kinematic suppression factor to account for the mass of the 
$X$ produced:
\begin{equation}
S_K = \left[ 1 - \left( \frac{m_X}{18.15\ {\rm MeV}} \right)^2 \right]^{3/2}.
\label{SK}
\end{equation}
One obtains thus the estimate:
\begin{equation}
\frac{|\kappa_p + \kappa_n|}{e}\ \sqrt{\br} \sim 1.16 \times 10^{-2}
\label{kappa1}
\end{equation}
or roughly:
\begin{equation}
\kappa_p \sim \kappa_n \sim 1.7 \times 10^{-3}/\sqrt{\br}.
\label{kappapn}
\end{equation}
In other words, when treating the $\kappa$ as parameters, there would be no 
difficulty in fitting the Atomki signal as it appears in experiment. 

The question, however, is whether this interpetation of the Atomki anomaly as 
$X$ in FSM would satisfy the contraints set by other experiments, which is 
always the main problem in interpetations of the Atomki anomaly, as so clearly 
analysed in the two excellent reviews \cite{Feng,Moretti} on the subject.  In 
Table \ref{constraints} are displayed the constraints listed in these reviews.

\begin{table}

\begin{tabular}{l|l|l|l|c}

No. & Experiment & Bound & Estimate & OK? \\ 
\hline 
$1^\#$ & M\o ller scattering & $C_{eV} C_{eA} < 10^{-8}$ & $\sim 0$ & OK \\
      & SLAC E158         &                        &     &    \\
\hline
$2^\#$ & Atomic parity     & $- \frac{2 \sqrt{2}}{G_F} C_{eA}$ & $\sim 0$ & OK \\
      & violation in      & $[C_{uV}(2Z+N)+C_{dV}(Z+2N)]$     &     &    \\
      & cesium            & $\times  \frac{0.8}{(17\ {\rm MeV})^2} < 0.7$ &    &    \\
\hline
$3^{*\#}$ & $\nu-e$ scattering & $|C_{e} C_{\nu}|^{1/2}< 10^{-4}$    & $\sim 0$ & OK \\
        & TEXONO             &  &     &    \\
\hline
$4^*$ & $\nu$-nucleus       & $|C_{\nu} C_{n}|^{1/2}< 2 \times
                              10^{-4}$
                         & $\sim 0$ & OK \\
      &  scattering         &   &     &    \\
\hline
$5^{*\#}$  & $e$ beam dump   & $|C_{e}| >6 \times 10^{-5} \times
                               \sqrt{\br}$  &  $\sim 3
                                                       \times 10^{-4}$
                                    & OK \\
         & SLAC E141       &  &                   & \\
\hline
$6^\#$ & $e$ beam dump      & $|C_{e}| > 1.3 \times 10^{-4} \times
                              \sqrt{\br}$
                         & $\sim 3 \times 10^{-4}$ & OK \\
      & CERN NA64         &   &                   &    \\
\hline
$7^*$ & $\eta \eta' \rightarrow \gamma (X \rightarrow e^+ e^-)$ &
        $|C_{e}| > 6 \times 10^{-6} \times \sqrt{\br} $ & $\sim 3 \times 10^{-4}$ & OK \\
     & CERN CHARM        &                    &                   &    \\
\hline
$8^{*\#}$ & $e^+ e^-$ formation & $|C_{e}|  < 6 \times 10^{-3}/\sqrt{\br}$ &
                               $\sim 3 \times 10^{-4}$ & OK \\
      & KLOE-2            &           &         &  \\
\hline
$9^{*\#}$ & $g - 2$ of electron & $|C_{e}| < 5 \times 10^{-4}$ &
                               $\sim 3 \times 10^{-4}$ & OK \\ 
\hline
$10^*$ & neutron-lead      & $|\kappa_{n}| < 6.0 \times 10^{-3}$
                           & $1.7 \times 10^{-3}/\sqrt{\br}$ & OK \\
       & scattering        &                               &     &    \\
\hline
$11^{*\#}$ & $\pi^0 \rightarrow \gamma (X \rightarrow e^+ e^-)$ &
                   (a) $\!|2C_{uV}\!+\!C_{dV}|  < 3.6\!\times\! 10^{-4}/\sqrt{\br} $
 & $\sim 3 \times 10^{-4}$  & OK \\
         & CERN NA48/2        & (b) $\!|\kappa_\pi|   <
                                10^{-3}/\sqrt{\br}$       & none known &?  \\
\hline

\end{tabular}

\caption{Constraints taken from the reviews \cite{Feng,Moretti}
  of Feng et al. (*) and Delle Rose et 
  al. (\#), in the notation of (\#) for the $X$'s couplings,
  where for simplicity we write $C_x=(C_{xV}^2+C_{xA}^2)^{1/2},
  x=e,\nu,n$, and \br\ stands for $\br(X \to e^+e^-)$.
  Since our 
  $X$ couples almost exclusively
  to leptons like a dark photon, it has practically no coupling to neutrinos and 
no $C_A$, so that 1---4 are automatically satisfied.  The fit to the Atomki 
signal gives no constraint on $C_{eV}$, so that one is free to choose $C_{eV} 
\sim 3 \times 10^{-4}$ to satisfy both the lower bounds 5, 6, 7 and the upper 
bounds 8, 9.  
For the 2 remaining constraints, 10 on $\kappa_n$ is satisfied
(conditionally---see text), 
while 11 is partly on (a) $|2C_{uV}+C_{dV}|$, which is satifsfied, and
partly on (b) $\kappa_\pi$ for which no estimate is yet known so no conclusion 
can yet be drawn.}

\label{constraints}

\end{table}

Note that in fitting the size of the Atomki signal only $|\kappa_p + \kappa_n|$ 
is involved; the coupling of $X$ to leptons is not constrained, which 
remains just dark photonic.  Thus:
\begin{itemize}
\item {\bf [n8]} $X$ acquires via mixing a component in $\gamma$, coupling 
thus to quarks and leptons like a dark photon, hence:
\begin{itemize}
\item (i) It can decay into $e^+ e^-$,
\item (ii) It has almost no coupling to $\nu$ and almost no axial coupling $C_A$
(which couplings would come only from the negligible $Z$ component), thus 
satisfying constraints 1---4 in Table \ref{constraints},
\item (iii) Its coupling to the electron is constrained by 5---9 in Table \ref{constraints} to 
have value $\sim 3 \times 10^{-4}$ which is small compared with $\kappa_{p,n}$ 
in (\ref{kappapn}), thus contributing but little to its production in beryllium 
decay.
\end{itemize}
\end{itemize}

However,
\begin{itemize}
\item {\bf [n9]} $X$ is not just a dark photon but a compound state bound by 
colour confinement.  Hence,
\begin{itemize}
\item it has additional compound couplings $\kappa_{p,n}$ which dominate $X$ 
production in $Be$ decay, 
\item yet, as will be shown immediately below, it satisfies constraint
  10 (conditionally) and 11 (a),  and 
by-passes constraint 11 (b) in Table \ref{constraints}.
\end{itemize}
\end{itemize}

Consider first the constraint 10 from neutron-lead scattering on $\kappa_n$.
Taking $\kappa_n \sim \kappa_p$ (since $C_p$ is small compared to $\kappa$),
one obtains from (\ref{kappa1})   the estimate $\kappa_n \sim 1.7
\times 10^{-3} / \sqrt{\br}$, which
is seen to satsify the constraint 10, provided $\br = \br (X \to e^+
e^-)$ is not too small, as it is believed not to be (see {\bf [R6]} in
the next section).
As for constraint 11, this is
more complex, receiving contributions as it does from both the dark photonic
and compound couplings of the $X$ to the pion.  The constraint (a) given in
\cite{Feng,Moretti} is obtained by putting the compound coupling
$\kappa_\pi$
of $X$ to the pion to $0$, which constraint is seen in Table 2 to be
satisfied.
Further, when the contributions of $\kappa_\pi$ is added, this
constraint may have to be weakened,
because of possible destructive interference between the two
contributions, and continues therefore to be satisfied.  A constraint (b) is
implied, however, also on $\kappa_\pi$, but no estimate for this is
yet known,
so no conclusion can as yet be drawn whether this constraint is satified or
not.
Estimates have indeed been obtained before for the compound couplings of $X$
to
nucleons $\kappa_{p,n}$.  But it is not obvious how a bound on $\kappa_\pi$
will translate (if at all) into bounds on $\kappa_p$ or $\kappa_n$, to
fathom
which one would have to understand the mechanism how the fundamental
couplings
of gluons, framons and quarks give rise to the effective couplings $\kappa$
connecting the compound state $G_3$ to the other compound states $p, n$ or
$\pi$.  Thus, this constraint 11 which is the most restrictive to other
models seems avoided by the present interpretation.

It is the fact that the $X$ in FSM is a composite state bound by colour 
confinement and so has a different interaction to hadrons than just that due
to the dark photonic coupling to quark constituents, thus separating the 
leptonic couplings $C_{eV}$ and so on from the hadronic couplings $\kappa_{p,n}$,
which allows the two sets of bounds 5-9 and 10-11 in Table \ref{constraints}
to be individually satisfied and keeps the present accommodation apparently
consistent.

In addition, this interpetation of the Atomki anomaly is seen to be consistent 
with the interpretation of the $g - 2$ and Lamb shift anomalies as $U$ in the
preceding section.  We recall that the $X$ boson here can in principle 
contribute to $g - 2$ and Lamb shift as well but was neglected there because, 
we argued, the couplings of $X$ to leptons being of 1-loop order should be 
small compared to the leptonic couplings of $U$ which occur already at tree 
level.  This checks with the numbers obtained above: 
\begin{itemize}
\item {\bf[n10]} The mixing parameter $\rho_{X \gamma}=C_e/e $ (that
  is, the $\gamma$-component of $X$)
 of order $10^{-3}$ is 
much smaller than $\rho_{Uh}$ of order $10^{-1}$, which is consistent with the 
FSM conception of $\rho_{Uh}$ being of tree-level and $\rho_{X \gamma}$ of 1-loop 
order.
\end{itemize}

Further, the couplings $\kappa_{p,n}$ of $X$ to nucleons should have equivalents 
for $U$ as well, both being composites bound by colour confinement, but in the 
analysis of the $g - 2$ and Lamb shift anomalies in Section 2, such couplings
were ignored, the couplings of $U$ to nucleons there being deduced only from 
the $U$ couplings to the constituent quarks.  Whence then the difference?  
Again, it is because the dark photonic couplings of $X$ are of 1-loop order 
while the dark Higgs-like couplings of the $U$ occur at tree level.  Hence, 
$C_{qV} \sim C_{eV} \sim 3 \times 10^{-4}$ are smaller than the compound 
couplings to the nucleons $\kappa_{p,n} \sim 1.7 \times 10^{-3}$ which dominate.
On the other hand, for $U$, it would be the other way round, with the dark 
Higgs-like couplings dominating over the compound couplings.  Hence the 
difference, namely,
\begin{itemize}
\item {\bf [n11]} The compound coupling $\kappa_{p,n}$ being of order $2 \times
10 ^{-3}$ is small compared with $g_{Up}$ of order $3 \times 10^{-2}$.  A 
parallel compound coupling for $U$ can thus reasonably be neglected at first 
instance, as it was in Section 2.
\end{itemize} 

Indeed, we think that compound couplings of $U$ to hadrons similar to the 
compound couplings $\kappa_{p,n}$ for $X$ very likely exist (See also {\bf [R4]}
of Section 4), making the $U$ couplings to hadrons more complicated than the 
way they were pictured in Section 2, as we already hinted at the end of that 
section.  This may not be too serious for the nucleon so that the analysis 
there still largely applies.  But when dealing with heavier nuclei, it is possible that 
the complication multiplies and helps further to explain why we had difficulty 
with Lamb shift for helium and with neutron-lead scattering.

\section{Concluding remarks}

It might have been premature trying to
accommodate experimental anomalies not yet fully 
established to a barely nascent theory with only limited predictive power but 
the attempt, it seems, has resulted in strengthened credibility and benefits 
for both.  We note in particular the following salient points.

\begin{itemize}
\item
{\bf [R1]} A first bizarre feature of the anomalies when considered as new 
particles is their abnormally small couplings to leptons, and probably also to 
quarks.  This fact has now found a possible explanation in the FSM in that the 
model predicts a hidden sector of particles with no direct coupling to 
those in the standard sector, but some of them, such as the scalar boson $H_+$ 
and the vector boson $G_3$, can acquire small couplings to quarks and leptons 
via mixing, the former with the standard Higgs boson $h_W$, and the latter with 
the photon.  As listed in {\bf [n1]}, {\bf [n8]}, and {\bf [n10]}, the fitted 
values of the various couplings are consistent with their being such mixing 
parameters.

Assigning the weakness of the anomalous signals to the smallness of mixing 
between a standard and a hidden sector, as the FSM does and as bourne out 
by the above analysis, has given the anomalies a very different significance to 
that obtained by assigning the same to the smallness of their couplings {\it 
per se}.  The latter would make of the anomalies a sort of appendix to the 
standard sector which one may choose to ignore at first instance.  But the 
former would make them both samples of and windows into a vast portion of our 
universe which has so far been hidden from us, a portion as complex and vibrant 
in itself as the portion we know and probably even weightier, as the amount of 
dark matter suggests.  And this we can hardly neglect.

On the other hand, although the existence of a sector of particles with no 
direct coupling to the standard ones, and that some among them, in particular 
$H_+$ and $G_3$, should acquire small couplings to quarks and leptons via 
mixing is a consequence of the FSM action, that the new sector should therefore 
be ``hidden'' depends on the assertion that these particles have none (or at 
least very little) of the soft interactions that hadrons have, despite the two 
types of particles having some similarity, both being compound states bound by 
colour confinement.  This conclusion was arrived at based on two parallel
arguments: 
\begin{itemize}
\item (i) by analogy \cite{fsmpop} to the Higgs scalar $h_W$ and the vector 
bosons $W^\pm, Z$ in the standard electroweak theory when this is pictured as 
a confined theory in the manner suggested by 't Hooft \cite{thooft}, 
\item (ii) by an intuitive argument \cite{cfsm} based on the observation that 
the framon constituents of the particles in the hidden sector have short 
lifetimes.
\end{itemize}
Neither the analogy nor the intuitive argument, however, need be considered conclusive.

Therefore, some phenomeological support for the validity 
of this ``hiddenness'' assertion in FSM would be welcome, and this seems now to be 
afforded by the anomalies as here interpreted.  Indeed, if the anomalies are 
real and are interpreted as particles, then these particles cannot have the 
strong soft interactions that hadrons have, or else they would have been 
produced copiously and recognized a long time ago.  Now, the only reminder one 
has of soft interaction in the above analysis is what we call the compound 
couplings $\kappa_{p,n}$ of the $X$ to nucleons in Section 3, and these we note 
are very small.  One can thus regard this fact as an experimental check on the 
validity of the above assertion that particles in the hidden sector have 
little soft interactions with particles in the standard sector.

\item
{\bf [R2]}  A second bizarre feature of the anomalies interpreted as particles 
is their mass of around 20 MeV, a mass region which does not play any
particular role in  anything else we 
know.  But it has found now an apparently natural home in the FSM, a model 
constructed originally, we recall, not for the anomalies but to address the 
generation puzzle in quarks and leptons.  Indeed, new particles in this mass 
range were predicted by the FSM before they were thought to have anything to do 
with the anomalies, as noted already in {\bf [n5]} and {\bf [n7]}, and this 
coincidence has added to their credibility.

However, this FSM prediction is itself in need of some scrutiny.  It was based 
on the following premises \cite{cfsm} 
\begin{itemize} 
\item (i) That the physical mass of a particle $x$ is to be given by a solution 
of the fixed point equation (\ref{mphys}), as applied above to $U$ and $X$, 
where $m_x(\mu)$ is its scale-dependent mass eigenvalue.
\item (ii) That $m_x(\mu)$ for the particles of present interest is as given by 
the fit to quark and lepton data reported in \cite{tfsm}, going sharply to zero 
at around 17 MeV as seen in Figure \ref{R}.
\item (iii) That if there are two solutions to (\ref{mphys}), then one takes 
the lower for the physical mass, the higher state being taken to be unstable 
against decay into the lower state.
\end{itemize}
Although (i) is generally accepted, and (ii) and (iii) already applied in the 
earlier fit \cite{tfsm} to quarks and leptons with apparent success, so that a 
mass prediction of aroung 20 MeV for $U$ and $X$ is no more than pushing the 
above to its logical conclusion, it is still an audacious conclusion to draw, 
predicting particles in this mass range in a system the natural scale for which 
would seem to be of order TeV.  But, as noted in {\bf [n5]} and  {\bf
  [n7]} in the preceding sections, experiments do seem to support this conclusion.
Thus, if both the anomalies and their present 
FSM interpetation are sustained, it would be a long shot which seems to have 
hit the mark and a major boost in the credibility of the logical process 
followed.  It suggests also the possibility of a deeper significance to the 
scale of around 20 MeV than is apparent at present \cite{Bjorken}.

\item
{\bf [R3]}  Next, we turn to details.  In the $g - 2$ and Lamb shift anomalies 
interpreted as due to $U$ boson exchange, we have noted in the $U$ couplings 
the following pattern:
\begin{itemize}
\item $g_{U\mu} > g_{Ue}$,
\item $g_{Up} \gg g_{U\mu}$,
\item $g_{Up} > g_{Un}$.
\end{itemize}
As noted already in {\bf [n2], [n2'], [n3], [n4]}, this pattern is very well 
explained by the FSM, lending support again to the validity of the result.

As far as the FSM is concerned, this pattern comes about from the Yukawa term 
(\ref{Yukawa}), which we have noted already is different from that of the 
standard model, and is thus something that one would like to check carefully 
against experiment.  The Yukawa coupling for FSM was constructed originally 
with the view of explaining the hierarchical mass and mixing patterns of quarks 
and leptons observed in experiment, which it did reasonably well \cite{tfsm}.
However, to achieve this, use has been made only 
of the value of the Yukawa coupling taken at the vacuum expectation value of 
the Higgs scalar (or flavour framon) field.  To test the model further, one 
would wish to ask whether this Yukawa coupling continues to hold even where the 
scalar field is away from, or fluctuates about, its vacuum expectation value.
This means comparing the couplings of the Higgs boson with other particles to 
experiment.  And this will be checking the implications of the Yukawa
term (\ref{Yukawa})
beyond the purpose for which it was originally constructed, or in other words, 
checking genuine predictions of the FSM scheme.

One example for such a check would be the decay rates of the Higgs boson $h_W$ 
to quark and lepton pairs \cite{Hdecay,btfit} which works out in FSM to be:
\begin{equation}
\Gamma(h_W \rightarrow \bar{f}_i f_i) \propto m_T^2 |{\bf f}_i \cdot \balpha|^4 
\label{FSMhdecay} 
\end{equation} 
where $\balpha$ is the rotating vector taken at the scale $\mu = m_h$, and 
${\bf f}_i$ is the state vector of the quark or lepton state $f_i$, both of 
which are vectors in 3-D generation space.  This prediction is very different 
from that of the standard model, namely $\Gamma(h \rightarrow \bar{f}_i f_i) 
\propto m_{f_i}^2$.  For the heaviest generation in each species the two models
differ little numerically in their predictions, since at high scales, the
rotation of $\balpha$ is slow and $m_T$ are just the masses of the heaviest
generations, and these predictions check well with experiment.  For the second 
generation, however, the FSM via (\ref{FSMhdecay}) predicts generally much 
lower rates compared to those predicted by the SM.  These differences can in 
principle be tested at the LHC.  Unfortunatley the two most favourable cases 
for rates: $h \rightarrow \bar{c}{c}, \bar{s}{s}$, are both heavily beset with 
background problems for detection, while the $\bar{\mu} \mu$ case which is not 
so beset has such a small rate that no events has yet been seen \cite{cfsm}.
For these reasons, these checks with Higgs decays are not as yet practicable.

However, we now note that the analyses of the $g - 2$ and Lamb shift anomalies 
in Section 2 have in fact supplied us with an equally detailed check on the 
Yukawa term (\ref{Yukawa}) to that afforded by Higgs decay, though at a much 
lower scale and admittedly less direct.  In accommodating the anomalies, one 
has already checked, as one has hoped to do in Higgs decay, the predictions of 
the FSM Yukawa term for the couplings of $h_W$ to leptons and quarks not only 
for different generations of the same species in {\bf [n2] [n2']} but also the 
couplings to quarks and leptons of different species in {\bf [n3], [n4]}.  And
here, besides, one has to run the vector $\balpha$ several orders of magnitude 
down in scale to order 20 MeV according to Figure \ref{R} obtained earlier in 
\cite{tfsm}, in effect making use of and testing the full armoury of the FSM 
prescriptions.  It is in fact quite surprising that the prescriptions seem 
still so far to hold good.

Had one adopted instead of (\ref{Yukawa}) the SM Yukawa term where the
couplings of $h_W$ to quarks and leptons of various species, and of various
generations of each species, are just proportional to the mass of that
particular quark or lepton, one would have obtained, 
for the $U$ couplings, still $g_{U\mu} > g_{Ue}$, but $g_{Uu} \ll g_{U\mu}$
and $g_{Uu} < g_{Ud}$, which in the same analysis would imply $g_{Up}
\ll g_{U\mu}$ and $g_{Up} < g_{Un} $, 
that is,  vastly at variance with the pattern noted above
from experiment.  But there is of course no reason for such a replacement. 
The SM has no call to explain the $U$ couplings since it predicts no
$U$ particle 
mixing with $h_W$, while in the FSM, the Yukawa term (\ref{Yukawa}) is part
and parcel of the development in \cite{cfsm,tfsm} which predicted the $U$ in
the first place, and cannot therefore be arbitrarily replaced.

\end{itemize}

Apart from the preceding examples of benefits going both ways, there are also 
some points where the above accommodation of the anomalies in
experiment sheds some more light on the FSM hidden sector and 
serves to settle some ambiguities in the theory.

\begin{itemize}
\item
{\bf [R4]} Two high mass states $H_{\rm high}$ and $H_{\rm low}$ thought earlier in 
\cite{cfsm} to be metastable and possibly detectable in experiment now appear 
in view of the analysis in Section 2 to be highly unstable against decay into 
$U$.

\item
{\bf [R5]} That $G_3$ or $X$ should have additional compound couplings $\kappa$ 
to hadrons other than those it acquires through its dark photonic couplings to 
quarks is a notion seemingly invented for the occasion to accommodate the 
Atomki anomaly in FSM, but actually it has already been envisaged before in 
\cite{cfsm} in a different context.  Because of the very short framon life-time,
 it was argued there that framon bound states like $G_3$ or $X$ would be 
point-like and reluctant to engage in soft interactions.  But it was also noted 
that this suppression of soft interactions is not absolute, not being due to a 
selection rule but only to a lack of probability.  Indeed, it was even suggested
 there (in Section 10) that residual soft interaction may be responsible for 
the decay of some charged states of high mass which are unwanted as remnants in 
the present universe.  Could then the compound couplings $\kappa$ introduced 
here in Section 3 be identified with the residual soft interactions envisaged 
there?  Both are concerned with the situation when such a framonic bound state, 
which otherwise appears pointlike, enters an environment as obtains inside a hadron 
where colour is deconfined.  Will such a state bound by colour confinement 
partly dissolve in such an environment, giving rise to the compound couplings $\kappa$
or the suggested residual soft interactions?  If so, then the unwanted
 high mass charged members of the hidden sector which worried us in 
\cite{cfsm} can decay by emitting, for example, a charged pion to a low mass 
and stable neutral member of the hidden sector, and be removed as remnants 
of our present universe.  Indeed, the value of $\kappa$ estimated by fitting 
the production rate of $X$ in Atomki in Section 3 would even give us then an 
estimate for their decay rate.  The question is certainly food for thought and 
may in future figure as an important ingredient in FSM, not just in the decay 
of the high mass charged states just mentioned, but also in the interaction 
rates of other particles in the hidden sector which may function as 
constituents of dark matter.  It might even suggest new ways of
detecting dark matter particles.

\item
{\bf [R6]} In the FSM, there is a colour analogue to the Yukawa coupling for 
the standard Higgs field (\ref{Yukawa}), an analogue not easy to
construct explicity, since the fermion field content is not yet completely
known.
There is thus a 
question whether some colour analogues of neutrinos, called co-neutrinos in 
\cite{cfsm}, could acquire a very low mass via a see-saw mechanism similar to 
that suggested for neutrinos, giving them a mass low enough for the 
$H_{\rm light}$ and $G_{\rm light}$ mentioned in the Introduction to decay into, 
making these bosons unstable and thus excluded as dark matter candidates. In 
view of the suggestion,  however, that $G_3$, a member of $G_{\rm light}$, is now 
identified as the Atomki anomaly $X$, this possibility becomes highly unlikely. 
The branching ratio $\br(X \rightarrow e^+ e^-)$, given the estimate 
(\ref{kappapn}), is bounded by the constraint 10 in Table \ref{constraints} to 
be not less than about 10 \%, while if there is a light co-neutrino, then the decay 
rate into a pair of these will be enormous in comparison, given that the 
coupling of $G_3$ to co-neutrinos will be the colour gauge coupling
$g_3$ \footnote{As a corollary, in the suggested absence of light
  co-neutrinos, $X$ will decay almost entirely into $e^+ e^-$ since
  it hardly couples to $\nu \bar{\nu}$ and is below $\mu^+ \mu^-$
  threshold, except perhaps for a small branching ratio into $\gamma
  \gamma$ (estimated in \cite{Moretti} to be about 1 \%)} .  
Therefore, if the identification of $G_3$ to $X$ is accepted, then there will 
likely be a bosonic component to dark matter as represented by the partners 
$G_{1,2}$ of $G_3$ in $G_{\rm light}$.   These $G_{1,2}$ have no mixing with the 
standard sector as $G_3$ has, and so cannot decay into $e^+ e^-$ as $X$ does, 
and having now also no light co-neutrinos to decay into, will seem stable and 
possible consituents of dark matter.  Similarly, so will the partners of $U$ or 
$H_+$ in $H_{\rm light}$. 
\end{itemize}

For these and other more obvious reasons, experimental confirmation of all the
3 anomalies are eagerly awaited.

\end{document}